\documentclass[11pt,twoside]{article}
\usepackage{asp2010}
\usepackage{hyperref}
\usepackage{graphicx}
\usepackage{natbib}

\resetcounters

\begin{document}

\title{Advances in understanding young high-mass stars using optical interferometry}
\author{Willem-Jan de Wit$^1$ 
\affil{$^1$European Southern Observatory, Santiago, Chile}}

\begin{abstract}
The closest examples of high-mass star birth occurs in deeply embedded 
environments at kiloparsec distances. Although much progress has been 
made, an observationally validated picture of the dominant processes which 
allows the central hydrostatic object to grow in mass has yet to be established. The observational
technique of optical interferometry has demonstrated its potential in the field of
high-mass star formation by delivering a milli-arcsecond infrared view on the complex 
accretion environment. We provide an overview of the scientific results obtained with 
multi-aperture telescope arrays and briefly discuss future instruments
and their anticipated impact on our understanding of massive young stellar objects. 

\end{abstract}

\section{Introduction}
Optical interferometry (OI) decouples the single telescope diffraction
limit from the angular resolution limit.  This can be achieved by
combining the light-beams coming from an array of telescopes in order 
to create interference fringes. OI as an
astronomical observing technique has come of age during the first
decade of this century. At the turn of the century, there were 
less than 100 refereed science papers that used OI data. This number 
grew exponentially to $\sim 500$ in the period between 2000 and 2011 
(Ridgway et al. 2012\nocite{2012IAUTB..28..292R}). 
General user facilities like the Very Large Telescope
Interferometer (Glindemann et al. 2004\nocite{2008SPIE.7013E..11H}),
the Keck Interferometer (Ragland et al. 2010\nocite{2010SPIE.7734E...1R}), 
CHARA (ten Brummelaar et al. 2011\nocite{2011arXiv1107.2890T}) 
and IOTA (Traub et al. 1998\nocite{1998SPIE.3350..848T}) have made this development possible. 
For the first time, a clear window on astrophysics is provided 
with milliarcsecond resolution in the optical and infrared wavelength
range.  Such angular resolution translates to direct access to sub-AU stellar
physics and probing the stellar surfaces of the largest stars in the sky. Additionally, 
the first OI synthesis images have provided new understanding of 
many challenging phenomena since 2007 (for a review, see 
Berger et al. 2012\nocite{2012arXiv1204.4363B}). Consequently, 
the technique  has opened new horizons in stellar and AGN physics.

The field of high-mass  star formation has been awarded with
significant progress thanks to increasing sensitivity of infrared (IR)
instruments and the exploitation of the IR wavelength region by
satellites like ISO, Spitzer, and Herschel. New views have been
obtained on outstanding questions regarding fragmentation and
formation of massive molecular cores from Giant Molecular Clouds,
radiative feedback during the main accretion phase, the physics of
molecular flows and the kinematic feedback on the natal molecular
clump, and the geometry of the accretion flow onto the young star
(Zinnecker \& York 2007\nocite{2007ARA&A..45..481Z}; Beuther
2011\nocite{2011BSRSL..80..200B}).  Many authors have noticed 
similarities between the accepted paradigm of (isolated) low-mass star
formation and the evidence for disks, jets and
outflows associated with young high-mass stars ($M_{*}>8\,{\rm M_{\odot}}$). 
Yet, the required high accretion rates, the high mass-infall
rates from the massive envelope, the feedback from the star, the
high-momentum mass outflows and last but not least, the massive star
multiplicity could, potentially, all spoil this similarity. These
concepts become manifest on short timescales and in a deeply embedded
environment. The closest high-mass SF regions are typically located at
kpc distances requiring sub-arcsecond resolution in order to advance
on these challenges. Such spatial resolution is what modern optical
and (sub-)mm interferometers provide.

The application of OI to the field  of massive star formation has been 
a relatively recent development (de Wit et al. 2007\nocite{2007ApJ...671L.169D}). 
The technical challenge of OI applied to the study of massive young stellar objects (MYSO) 
is to collect enough {\it optical and near-IR} photons in order to stabilize (in real-time!)
the various telescopic and interferometric control loops. This poses challenges to applying
the technique to young massive stars. MYSO are optically dark, near-IR faint and
only in the mid-IR the sources become detectable. Added to this is 
the complication that not only the science target but in general the large majority 
of the neighbouring stars are buried in the natal cloud core and therefore relatively 
faint at optical and near-IR wavelengths.  In case of the VLTI, the 
adaptive optics system MACAO, the fringe-track system FINITO, and the 
IR tip-tilt system IRIS work with optical or near-IR photons. 

This brief review aims to fit the results obtained with OI into our growing understanding of
the birth of individual high-mass stars. We give a compact overview of 
the advancements over the past years before detailing the contributions by OI.

\section{Optical interferometry at the VLTI}
The only interferometric array that has achieved scientifically validated 
OI-fringe data  on MYSOs so far is the VLTI array, operated by ESO 
on Cerro Paranal. The VLTI is a versatile 
interferometric array that schematically consists of three basic
elements: 4 {\it telescopes} that feed an interferometric {\it
  beam-combiner instrument} after the optical path length is equalized via a
{\it delay-line carriage system}. The employed telescopes can
either be the four 8.2m Unit Telescopes or the four mobile 1.8m
Auxiliary Telescopes (see Hagenauer et al. 2008).  The
telescope platform on Paranal is equipped with 30 possible AT
locations (``stations'') which are connected with the delay-line
tunnels by underground light ducts. In operation mode, the VLTI offers
$\sim 9$ stations which are subdivided into three distinct AT
quadruplet configurations.  The baseline lengths of the currently
offered quadruplets range between 11 and 140 meters. Since 2005, ESO
has offered to the community 2 single-feed, spectro-interferometric, instruments. AMBER operates
in the near-IR. It combines 3 telescope beams delivering spectrally 
dispersed interferometric fringes for three spectral resolutions
(Petrov et al. 2007\nocite{2007A&A...464....1P}). MIDI is a 2
telescope combiner and delivers fringes at two possible
spectral resolutions in the N-band (Leihnert et
al. 2003\nocite{2003Ap&SS.286...73L}). The combination of wavelength 
and baseline results in a resolution of $\lambda/2B=1.6$\,milli-arcsecond in K-band.

Spectro-interferometry is the technique which allows the combination of high spectral with
spatial resolution. It is a powerful tool to perform spatially
resolved kinematic studies of, for example, gaseous circumstellar
disks (see e.g. {\v S}tefl et al. these proceedings).  By means of
fringe phase measurements, the VLTI data give access to photocentre
displacement and asymmetries on angular scales
of tens of micro-arcseconds (e.g. Wheelwright et al. 2012b).
For reference, the angular resolution of ALMA is 5 milli-arcsecond at
the shortest wavelength (0.3 mm) and largest configuration
(14.5 km), similar to the angular resolution offered by
the VLTI.  This complementarity is of interest to massive
star formation studies in, for example, uniting hot accretion physics
with the cold disk physics aiming for a complete picture of a massive
star accretion disk. Last year, 2011, marked the 10th
anniversary of the VLTI ``first light'' with a dedicated ESO science
meeting. All presentations are available on the
internet\footnote{\url{http://www.eso.org/sci/meetings/2011/VLTI_2011/program.html}}.

\section{Boundary conditions of high-mass star formation}
Star formation occurs in molecular clouds, but clearly not all
molecular clouds produce high-mass stars. Clouds that do are a subset
of the Infra-Red Dark Clouds (IRDCs, Egan et
al. 1998\nocite{1998ApJ...494L.199E}; Kauffman \& Pillai
2010\nocite{2010ApJ...723L...7K}; Peretto \& Fuller
2010\nocite{2010ApJ...723..555P}). These are cold and dusty
filamentary interstellar structures with high column densities which
renders them optically thick to the background mid-IR radiation field.
These physical conditions are thought to be representative of the
boundary conditions for the formation of high-mass molecular cores:
the analogue of the low-mass starless cores.  High-mass cores are 
envisaged to collapse without any significant fragmentation (i.e mono-lithic), resulting
in an individual high-mass star (McKee \& Tan 2003\nocite{2003ApJ...585..850M}).  However, such high-mass cores have 
escaped unambiguous detection (e.g. Caselli 2011\nocite{2011IAUS..280...19C}),
although angular resolution and sensitivity in the mm wavelength regime
may be responsible for the current non-detection (see Krumholz, these
proceedings). On the other hand, a real absence of massive pre-stellar cores would imply that
molecular cloud clumps fragment down to low-masses. How would such a scenario give rise
to a stellar mass distribution extending up to high-mass stars? One
idea envisages that the fragments' motion through the molecular
clump results in different (Bondi-Hoyle) accretion rates and high-mass
objects would materialize because the progenitor fragment happened to
accrete material from the densest clump regions (i.e. nominally the clump centre). 
A recent overview containing a detailed description of both models can be found in
Bonnell \& Smith (2011)\nocite{2011IAUS..270...57B}. For this review it suffices to state
that this earliest, pre-stellar, phase of high-mass stars does not
have any clear-cut observational counterpart, and are therefore
observationally inaccessible. A large body of work is directed towards
finding these objects with e.g. Herschel (Motte et al. 2010\nocite{2010A&A...518L..77M};
Molinari et al. 2010\nocite{2010A&A...518L.100M}) and ALMA 
will have a crucial role to play.

Nonetheless, somewhat more advanced phases in the high-mass star
formation process within IRDCs are easily identifiable. Deeply buried
by the molecular core material, a central object warms its immediate
surroundings up to several 100s of degrees. The elevated temperatures
are betrayed by molecular line transitions: a variety of chemical
species, initially frozen onto the dust grains, sublimate and are
released into the IS medium.  These species give rise to the typical
``hot core'' spectrum characterized by a plethora of (sub-)mm line
transitions of different complex organic molecules (Cesaroni et al. 2005,
2010\nocite{2005IAUS..227...59C}\nocite{2010A&A...509A..50C}). Additionally,
these early phases of a young high mass star are characterized by
compact and bright mid-IR emission (Henning et
al. 1990\nocite{1990A&A...227..542H}), bipolar, high-velocity CO
emission (Snell et al. 1988\nocite{1988ApJ...325..853S}; Beuther et
al. 2002\nocite{2002A&A...383..892B}), shock released SiO emission
(Jim\'{e}nez-Serra et al. 2010\nocite{2010MNRAS.406..187J}), and
water/methanol (6.7GHz) maser emission (e.g. Walsh et
al. 1997\nocite{1997MNRAS.291..261W}).  These are all hallmarks of
young high-mass stars ($L_{\rm bol}>10^{4}\,{\rm L_{\odot}}$), and
this embedded phase is generally identified as the massive YSO
(Lumsden et al. 2002\nocite{2002MNRAS.336..621L}), accessible to
detailed study.


\begin{figure}[!t]
    \includegraphics[clip=true,width=7.0cm,height=6.0cm]{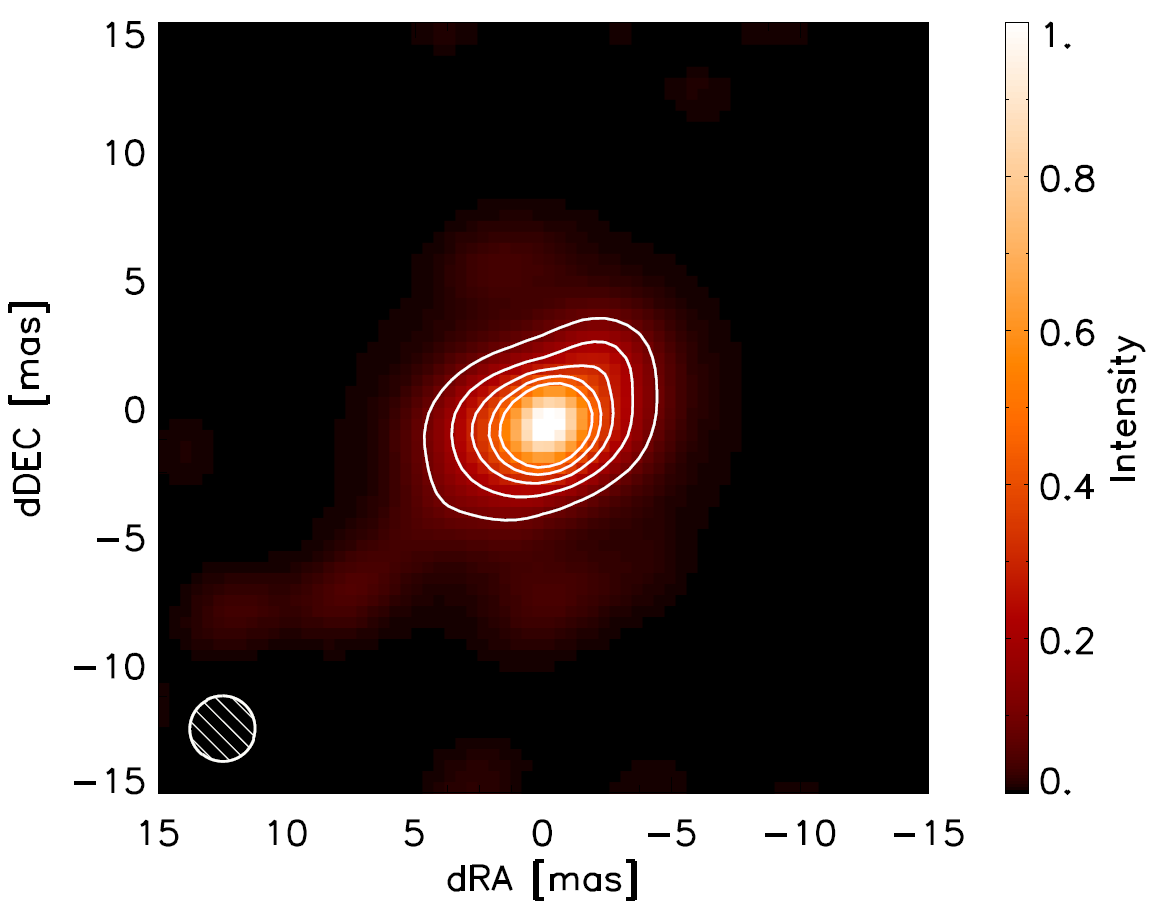}
   \includegraphics[clip=true,width=5.5cm,height=5.8cm]{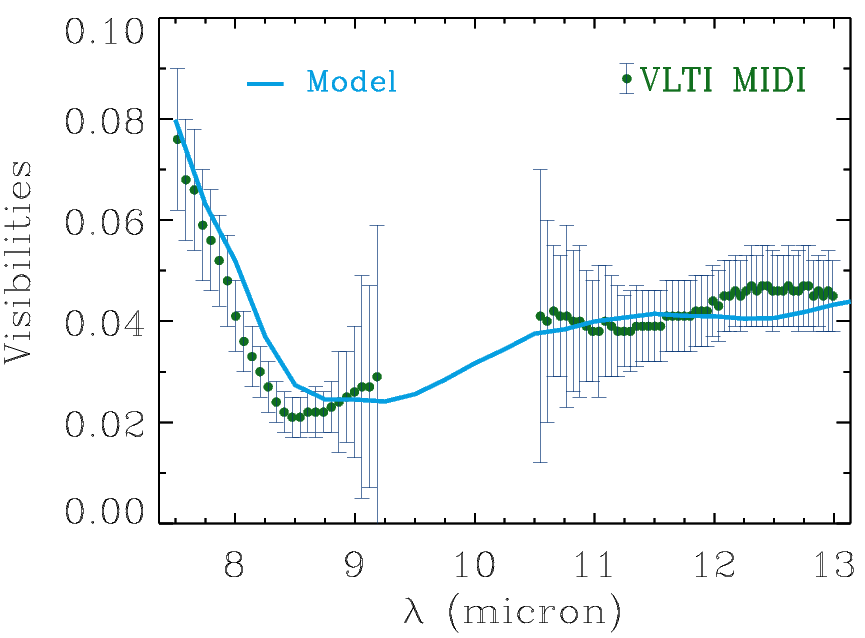}
    \caption{{\it Left:}  AMBER OI synthesis image by Kraus et al. (2010) (Reprinted by permission from Macmillan Publishers Ltd:  
      \href{http://www.nature.com/nature/journal/v466/n7304/full/nature09174.html}{Nature 466, 2010})  {\it Right:} Dispersed MIDI 
      visibilities (with errorbars) and model (full line) which combines a gaseous accretion disk and dusty envelope (de Wit et al. 2011).}
\end{figure}

\section{Massive YSOs and accretion}
One of the main questions in high-mass SF regards the accretion of mass
by the central, embedded, star. MYSOs display a number of
observational properties which are consistent with (very) high
accretion rates, where careful estimates based on molecular outflows
seem to indicate values up of $\rm 10^{-3}\,M_{\odot}\,yr^{-1}$
(Beuther et al. 2002\nocite{2002A&A...383..892B}; Zhang
2005\nocite{2005IAUS..227..135Z}).  The most recent accretion disk
simulations show that high accretion rates can be sustained by means
of very effective gravitational torques leading to non-axisymmetric
distribution of the disk material (Kuiper et
al. 2011\nocite{2011ApJ...732...20K}). The presence of a disk
significantly reduces the adverse effects of the radiation pressure on
the infalling material by preferentially beaming photons into the
polar direction (Krumholz et
al. 2005\nocite{2005ApJ...618L..33K}). Until recently, direct
observational evidence (i.e. a spatially resolved image) of an
accretion disk around a young high-mass star was impossible because of
the angular scales involved. The best evidence until recently was
probably the 130\,AU disk detected in mm continuum near a $\rm \sim
10\,M_{\odot}$ star (Shepherd et
al. 2001\nocite{2001Sci...292.1513S}). This situation changed
radically with the creation of a VLTI interferometric synthesis image
in the near-IR (Fig.\,1). From AMBER visibilities and closure phase
measurements supplemented with short spacing, NTT speckle images,
Kraus et al. (2010\nocite{2010Natur.466..339K}) applied image reconstruction techniques
to create a synthesis image of the MYSO IRAS +13481 ($\rm \sim 20\,M_{\odot}$)
with a spatial resolution of 2.4 milli-arcseconds ($\sim$8.4AU).  This
is the highest resolution MYSO image ever created to date. The image
shows a structure with a smooth spatial profile that is centrally
peaked. It strongly suggests a disk geometry for the continuum
emitting material. This interpretation is re-inforced by the fact that
the disk's orientation is perpendicular to that of a large-scale,
bipolar, CO outflow. In addition, this outflow may be driven by an
accretion induced jet which is the main suspect of a parsec-scale
collimated H$_2$ flow (Stecklum et al. 2010). (The Br$\gamma$
emission is scrutinized in sub milli-arcsecond detail by SINFONI
integral field spectroscopy, see Stecklum et al. these proceedings.)


The reconstruction of a near-IR image at a resolution of 2.4 milli-arcsecond
constitutes an important step forward in understanding the
circumstellar environment of an accreting high-mass star, yet the
interpretation of the OI synthesis image is not straightforward. The
observed spatial distribution of the near-IR emission together with
the source's SED are consistent with 
reprocessed stellar emission by a dusty disk+envelope system. In
this particular model, the disk mass is the same as that of the central
stellar source, i.e. $\rm \sim 20\,M_{\odot}$. The inner disk is
located at a radius of 6.2\,AU, which roughly corresponds to the dust
sublimation radius.  As a consequence, IRAS +13481 appears to comply with
the well-known near-IR luminosity-size relation established for lower
mass pre-main sequence stars (see e.g. Dullemond \& Monnier 
2010\nocite{2010ARA&A..48..205D}). 
If this interpretation is correct, a very attractive
unifying scenario would emerge in which the AU-scale, near-IR emission of
all accreting young stars is dominated by dust sublimation
physics. Indeed, the AMBER synthesis image is found to be consistent with 
a warm, puffed up inner disk rim.  However, confrontation with a $20\,\mu$m single-dish 
image proves the limitations of the particular envelope model (Wheelwright et al. 2012a), 
while observations on milli-arcsecond spatial scales provided by the VLTI-MIDI instrument 
also shows that a puffed-up dust sublimation rim as the origin for a large fraction 
of the near-IR photons has some trouble and may need revision  (Boley et al. in prep.).

The OI technique offers the unique possibility to study the morphology of 
circumstellar disks around young high-mass stars at wavelengths longer 
than $K$-band. At the near-IR wavelengths, 
scattering and envelope extinction hamper the direct detection of disk light. At 
longer, mid-IR wavelengths, the contribution by the warm envelope dust ramps
up, slowly drowning disk photons. There is a 'sweet spot' in terms
of wavelength between K and N-band where the circumstellar disk is
prominent and may dominate the total light. Interferometry of MYSOs
using MIDI in the $N$-band probing $\rm \sim 100\,AU$s clearly 
demonstrates the absence of rotational symmetry in a couple of case studies. Subsequent
radiative transfer modelling suggests that the geometry of the
$N$-band emission is consistent with that of an equatorially flattened
structure (Follert et al. 2010\nocite{2010A&A...522A..17F}; Grellmann et
al. 2011\nocite{2011A&A...532A.109G}).  The power of interferometry is that the various
contributions to the correlated fluxes is the flux weighed by the
size; at milli-arcsecond resolution the envelope emission is almost
resolved out, and any contribution by an unresolved structure e.g. a
disk will become evident. Such an interpretation of MIDI observations
of the well-known embedded young massive star AFGL\,2136 is presented
by de Wit et al. (2011\nocite{2011A&A...526L...5D}). An increase in visibilities at the blue edge
of the N-band (see Fig. 1) is consistent with the presence of an alpha-type
accretion disk located within ($\rm <170\,AU$) the dusty envelope,
accreting at a rate of $\rm 3 \times 10^{-3}\,M_{\odot}\,yr^{-1}$.  A
remarkable detail, imposed by $N$-band, short-spacing Keck data
(Monnier et al. 2009\nocite{2009ApJ...700..491M}), is that the rim of
the dust envelope is found at about 7 times the formal dust
sublimation radius. The central dust-free zone could have been
evacuated by the ionized stellar wind, seen by Menten \& van der Tak
(2004\nocite{2004A&A...414..289M}).

It is clear that both in-depth OI studies to clarify the accretion physics and 
survey-type of OI work to probe the frequency of disks and their properties 
are badly needed to provide a full picture of MYSOs as function of e.g. 
luminosity or age.

\section{Observational consequences of disk accretion}

\begin{figure}[!t]
    \includegraphics[clip=true,width=13.0cm,height=5.5cm]{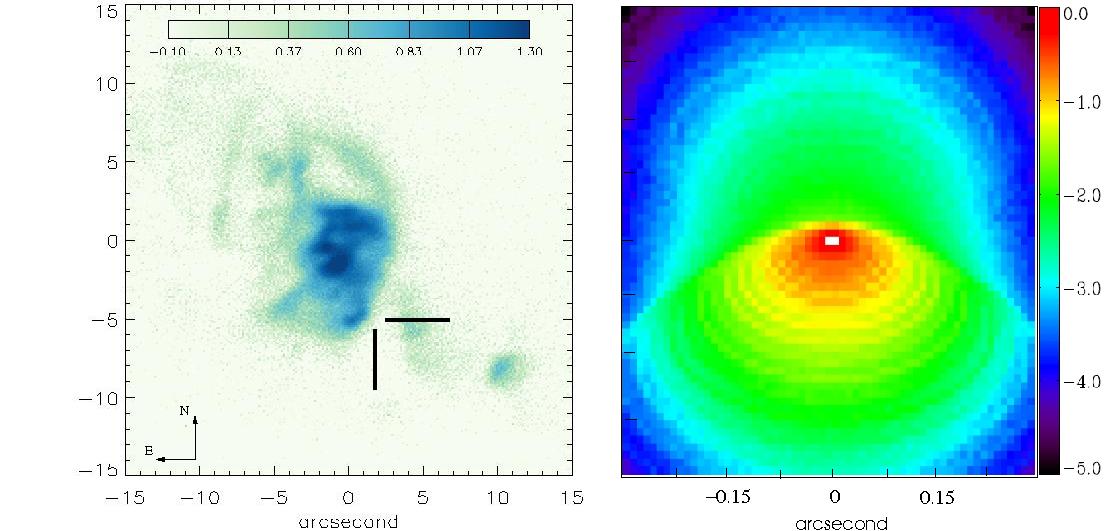}
    \caption{{\it Left:} Large scale thermal cavity wall emission at $24.5\,\mu$m in Cep\,A HW2. The two bars indicate the phase centre 
    (i.e. the central source) of the VLA 7mm map by Jim\'{e}nez-Serra at el. (2007), adapted from de Wit et al. (2009). {\it Right:} RT
model image at $12.5\,\mu$m of an MYSO envelope with evacuated cavities which fits MIDI visibilities of  W33A. The image is 0.6\arcsec \,on each side  (de Wit et al. 2010).}
\end{figure}

The presence of small-scale disks around MYSOs is inferred from the upper limits of high-precision 
spectro-astrometric signals of near-IR CO bandhead emission (Wheelwright et al. 
2010\nocite{2010MNRAS.408.1840W}). Performing OI and spectrally resolving 
near-IR spectral features like the CO-bandhead or the hydrogen recombination 
lines of MYSOs is currently at the sensitivity limit of the VLTI. It can be expected to shed light 
on the geometry and origen of the warm molecular material and the hot gas in the near future.
It is important to keep in mind that the envelopes of MYSOs can be
optically thick even in the mid-IR (e.g. Fig. 2) and the ability to peer deep into the 
inner regions depends critically on the inclination and the evolutionary stage. Radio 
emission is thus highly suitable to study the most extincted regions.  Unfortunately, 
MYSOs display little ionized emission despite their high $L_{\rm bol}$.  Any weak radio 
emission is found to be compact ($\sim 100$ AU, van der Tak \& 
Menten 2005\nocite{2005A&A...437..947V}). The nature of this emission
is under debate but it could be connected to the outflow activity.
For example, Guzm\'{a}n et al. (2010\nocite{2010ApJ...725..734G}) report the discovery of a
string of radio blobs emanating from either side of  a $7\,10^{4}\,{\rm L_{\odot}}$ source.  
A highly collimated radio jet is present in Cep\,A HW2 (Torrelles et al. 1996\nocite{1996ApJ...457L.107T}) 
displaying velocities of $\rm 500\,km\,s^{-1}$  (Curiel et al. (2006\nocite{2006ApJ...638..878C}). 
Spectro-astrometry of the MYSO W33A, probing photo-centre shifts of $\sim 100\,\mu$-arcsecond, already shows a 
bipolar geometry for the Br$\gamma$ emission with a position angle that coincides with 
the arcminute molecular outflow (Davies et al. 2010\nocite{2010MNRAS.402.1504D}). 

The ubiquity of CO outflows in MSF regions and their scaled up
properties compared to those of less luminous systems is indicative of
a common mechanism (Shepherd et al 1998\nocite{1998ApJ...507..861S}; 
Beuther et al. 2002).  Varricatt et al. (2010\nocite{2010MNRAS.404..661V}) probe the 
incidence of linear, shock excited molecular H$_{2}$ emission at $\rm 2.12\,\mu
m$ associated with MYSOs. The authors address the question whether jets are the driving
mechanism of CO outflows (Torrelles et al. 2011\nocite{2011MNRAS.410..627T}) rather than
the wind of the underlying star. Approximately 50\% exhibit H$_{2}$
emission features along the molecular outflow direction in a sample
of 50 massive SF regions. They also find that the H$_{2}$ emission is absent
or comparatively weak towards UCH\,II regions. This is in line with the idea that 
high mass infall rates quench the formation of an H\,II region (Walmsley 1995\nocite{1995RMxAC...1..137W}); the
absence of radio emission is a proxy for ongoing accretion. 
Infall rates determined from sub-mm studies of rotating motion on scales of $\rm > 1000\,AU$  
indicate infall rates even higher than those determined from molecular outflows: 
 $\rm 10^{-3}-10^{-2}\,M_{\odot}\,yr^{-1}$  (Beltran et al. 2011\nocite{2011A&A...525A.151B}). 

Sustained high accretion rates (possibly episodic as deduced from $\rm H_{2}$ flow
morphology) induce a considerable effect on the stellar structure of the
growing, quasi-hydrostatic, stellar core. With time, the
Kelvin-Helmholtz timescale will decrease non-linearly and becomes
shorter than the star's evolutionary timescale. The stellar interior will
effectively transport radiation outward because of the decreasing Kramer's
opacity, resulting in a swelling of the protostar (Hosokawa, York \& Omukai 2010\nocite{2010ApJ...721..478H}).  OI with the MIDI instrument 
has provided some preliminary evidence that a swollen star is present in the source M8E-IR (Linz et al. 2009). 
The predicted stellar radius of these rapidly accreting object ($\rm \sim
100\,R_{\odot}$) will strongly reduce the accretion
luminosity associated with the disk ($L_{\rm acc}\propto R_{*}^{-1}$) and thus the 
observed bolometric luminosity should be dominated by emission from the central
object. Importantly, the swollen star mechanism constitutes a noteworthy alternative
for the faint, compact radio emission of MYSOs. Rather than the ionized emission 
being quenched by high accretion rates, the photosphere of the
central object is cool and does not emit any ionizing radiation. The observed compact emission could be
caused by shocks generated by the interaction of the collimated wind with the
surrounding envelope material and cavity walls (e.g. Parkin et al. 2009\nocite{2009MNRAS.400..629P}).

\section{Milli-arcsecond observation of the outflow cavity}
Spherical geometries for the MYSO envelope generally fail to reproduce the observed fluxes at   
wavelengths shortward of $20\,\mu$m (e.g. Mueller et al. 2002\nocite{2002ApJS..143..469M};
de Wit et al. 2009\nocite{2009A&A...494..157D}). Such models are unable to predict correctly 
the depth of the $10\,\mu$m absorption feature and the overall shape of the (sub-)mm SED. 
Attempts to shed light on this issue by means of spatially resolved mid-IR imaging failed 
when using 4m apertures (e.g. Mottram et al. 2007\nocite{2007A&A...476.1019M}). 
Observations with 8m class telescopes have been able to spatially
resolve the mid-IR emission for a couple of case studies. They 
revealed that the mid-IR emission 
is aligned with the larger scale  
CO molecular outflows (De Buizer 2006, 2007\nocite{2006ApJ...642L..57D}\nocite{2007ApJ...654L.147D}). 
Segment tilting interferometry at N-band with the 10m Keck also resolves 
the MYSO envelope demonstrating complicated, often bipolar, structures 
(Monnier et al.  2009\nocite{2009ApJ...700..491M}). These findings 
associate the mid-IR emission of massive YSOs with outflow cavities, or,
more specifically, with parts of the dense, dusty envelope close to the 
evacuated cavities which experience near-direct irradiation by the
hot central parts.
 
N-band OI performed with MIDI on unresolved MYSO sources  
spatially resolves the emission on scales of tens of milli-arcseconds or
 $\sim 100$\,AU.  Radiative transfer modelling allowed the conclusion
that the emission is consistent with thermal envelope emission (de Wit et
al. 2007\nocite{2007ApJ...671L.169D}; Linz et al. 2009\nocite{2009A&A...505..655L}).  In particular, a
detailed modelling is presented in de Wit et al. (2010\nocite{2010A&A...515A..45D}) for
the luminous MYSO W33A.  The authors manage to model the MIDI
interferometry, near-IR scattered light images and the SED simultaneously.
A dust geometry was chosen which includes a rotating, infalling
envelope with dust-devoid, polar paraboloids representing the cavities
carved by the jet/outflow. An excellent fit was found to
all observables and which, moreover reproduces the observed 350 micron
morphology (van der Tak et al. 2000\nocite{2000ApJ...537..283V}) quite accurately. Close
inspection of the envelope temperature distribution clearly reveals
that the N-band emission is completely dominated by
warm dust in the cavity walls on 100\,AU scales, directly irradiated by the central
star (see Fig. 2). The accretion rate of a circumstellar $\alpha$-type
disk would need to be more than $\rm 10^{-3}\,M_{\odot}\,yr^{-1}$ for it to exceed
the cavity wall emission.

The success of this model was recently extended in Wheelwright
et al. (2012a\nocite{2012A&A...540A..89W}).  The authors succesfully apply 
this model to simultaneously recreate the  resolved Q-band emission and SEDs for
a large MYSO sample. They simply vary the cavity opening angle, inclination and 
the total amount of dust of the W33A model. The mere fact that 
this approach works quite well argues  against a large effect, if any by a
circumstellar disk on the Q-band emission.  This conclusion can also be
inferred from an accretion disk SED, since it peaks at wavelengths shorter than 20\,$\mu$m. 
Furthermore, the envelope contains much more cool material
than does any reasonable accretion disk, whose sizes for MYSOs are
estimated to be less than 500\,AU (see e.g. Patel et al. 2005\nocite{2005Natur.437..109P}; Hoare
2006\nocite{2006ApJ...649..856H}; Reid et al.  2007\nocite{2007ApJ...664..950R}; 
Jim\'{e}nez-Serra et al. 2007\nocite{2007ApJ...661L.187J}; 
de Wit et al. 2011\nocite{2011A&A...526L...5D}; 
Goddi et al.  2011\nocite{2011ApJ...728...15G}). It is reasonable to
expect that dust emission from the envelope (i.e. the cavity walls)
dominates the continuum SED at wavelengths longward of the
N-band. This is in agreement with the finding that the observed
extension of the Q-band flux can be modelled as cavity wall emission.
Resolved imaging and especially OI thus deliver excellent possibilities 
to probe the base of the outflow and explore in detail the mass 
loss processes during the accretion phase.

\section{Summary and outlook}
Observational and numerical evidence is accumulating that accretion
disks can build stars up to $\rm 25\,M_{\odot}$. Although no
observational evidence exists that stars of higher masses follow the
same channel, numerical simulations clearly allow for the accretion disk
mediated formation of stars of  $\rm 100\,M_{\odot}$ and more.  
The disk density distribution and the high mass infall rates will lead 
to fragmentation and collapse of secondary and multiple objects 
(Krumholz et al. 2009\nocite{2009Sci...323..754K}) and the disk is expected 
to remain relatively small (Kratter et al. 2008\nocite{2008ApJ...681..375K}). OI has provided 
the first direct evidence for a disk near a high-mass star at a spatial 
resolution of 8\,AU in the near-IR. The increased sensitivity of 
the VLTI, implementation of new sub-systems and observing modes,
and sophisticated post-processing algorithms will potentially make spectrally 
resolved studies of ionized gas and warm molecular emission possible 
on similar spatial scales.

The warm dust located close to the outflow cavities is responsible for
a large fraction of the mid-IR emission emanating from structures on
100\,AU scales.  In hydrodynamic simulations of MYSOs, the
stellar wind momentum is not sufficient to break out of the infalling
envelope by an order of magnitude or more and would lead to
Rayleigh-Taylor instabilities resulting in the collapse of the
cavities (Krumholz et al. 2005\nocite{2005ApJ...618L..33K}; see also Kuiper et al
2012\nocite{2012A&A...537A.122K}). Seifried et al. (2012\nocite{2012MNRAS.422..347S})
confirm the appearance of a fast jet generated by a protostellar disk,
but whether a fast jet is generated or a slow wide-angle outflow
depends critically on the magnetic field strength.  Tracing the jet
structure gives a unique insight into the mass accretion process
itself (e.g. Ellerbroek et al. these proceedings).  A 3D magnetic
field morphology of the disk-jet system Cep\,A HW2 could be reconstructed
via maser polarization measurement using VLBI (Vlemmings et
al. 2010\nocite{2010MNRAS.404..134V}) demonstrating that the magnetic
field may play a crucial role in the formation of a MYSO jet
(Carrasco-Gonz\'{a}lez et al. 2010\nocite{2010Sci...330.1209C}).  
We have detailed how OI delivers complementary spatial information 
in order to probe these various ideas and to advance on our 
understanding of the accretion/outflow mechanism in MYSOs.

The coming decade will see the VLTI as the only large aperture interferometric array. 
The second generation VLTI instruments will offer more sensitive and more efficient 
interferometric observations with Gravity and MATISSE being both 4-beam combining instruments. 
Especially MATISSE (first light 2015) offers a strong potential for 
permanent breakthroughs in our understanding of the formation of individual 
high-mass stars. This instrument will open up the wavelengths between $K$ and $N$-band 
for OI, perfectly fitting for detailed studies of MYSO accretion physics. 


\acknowledgements 
I would like to thank the organization who provided me with the opportunity of presenting these results.
Furthermore, I am grateful to all collaborators who have contributed to our common efforts to understand
MYSOs. This manuscript was improved by the constructive comments supplied by R.D. Oudmaijer, 
B. Stecklum, J.S. Vink and H. Wheelwright.
\vspace{0.3cm}

\hspace{-0.4cm}{\bf Questions:}
\vspace{0.1cm}

\noindent{\bf C. Martayan:} There are 25-30$M_{\odot}$ objects in your presentation. How did you do these estimates? What
should be the resulting stellar masses of these objects?\\
{\bf W.J. de Wit:} The mass is a direct conversion from the $\rm L_{bol}$ and ZAMS values. MYSOs with higher
masses are not known. Indeed there is evidence that for the inferred $\dot{M}_{\rm acc}$  the hydrostatic object reaches
the ZAMS at $\sim 30\,M_{\odot}$ and will ionize its surroundings (see e.g. Davies et al. 2011\nocite{2011MNRAS.416..972D}).
\vspace{1.5mm}

\noindent{\bf J. Bjorkman:} How well can you constrain the cavity opening angle from the N-band visibilities?\\
{\bf W.J. de Wit:} Not at all. That's why we model the near-IR scattering nebula  in case of W33A 
for example in order to obtain a handle on the outflow opening angle.

\bibliography{dewit_foz_v3.1}

\end{document}